\documentclass[12pt]{article}
\usepackage[a4paper,margin=1in]{geometry}
\usepackage{amsmath,amssymb,bm}
\usepackage{graphicx}
\usepackage{slashed}
\usepackage{dcolumn}
\usepackage{algorithm}
\usepackage{algpseudocode}
\usepackage{multirow}
\usepackage{titlesec}
\usepackage{newtxtext,newtxmath}
\usepackage{dsfont}
\usepackage{xcolor}
\usepackage{mathbbol}
\usepackage{cite}
\usepackage{caption}
\usepackage{hyperref}
\hypersetup{
    colorlinks=true,
    linkcolor=blue,
    citecolor=blue,     
    urlcolor=blue,
}

\allowdisplaybreaks

\usepackage{mathrsfs}
\makeatletter

\newcommand{\Rmnum}[1]{\expandafter\@slowromancap\romannumeral #1@}
\makeatother

\title{Quantum phase-field model: vortices and THz-induced gap dynamics in superconductors}

\author{
  R.~Y.~Fang$^{1}$,~~  
  F.~Yang$^{2}$\thanks{Email: fzy5099@psu.edu},~~
  S.~L.~Zhang$^{1}$,~~
  L.~Q.~Chen$^{1,2}$\thanks{Email: lqc3@psu.edu}
}

\date{} 

\begin{document}

\maketitle

\vskip -5cm
\begin{center}
$^{1}$Department of Engineering Science and Mechanics, The Pennsylvania State University,\\
University Park, PA 16802, USA\\[4pt]
$^{2}$Department of Materials Science and Engineering and Materials Research Institute,\\
The Pennsylvania State University, University Park, PA 16802, USA
\end{center}

\begin{abstract}

The ability to simulate the spatial and temporal ordering dynamics of quantum phases in inhomogeneous systems, particularly in the low-temperature regime, is crucial for understanding condensate physics and for enabling emerging device technologies. However, fully microscopic kinetic  approaches are often computationally prohibitive for macroscopic spatiotemporal simulations in realistic device geometries, while the widely used phenomenological Ginzburg–Landau formulation is, in principle, valid only near the critical transition temperature $T_c$ and becomes inaccurate in the technologically relevant low-temperature regime. To describe the dynamics of quantum phase ordering,  we propose a phase-field formulation derived from a microscopic many-body description using the superconducting order as an example. It leads to a compact dynamical evolution equation for the superconducting order parameter, enabling real-space and real-time simulations of phase ordering dynamics.  The simulations successfully capture key dynamical phenomena, including vortex nucleation and motion under static magnetic fields, as well as ultrafast gap oscillations driven by THz fields,  over the full temperature range from 0~K up to the critical transition temperature. Beyond superconductivity, this method can be extended to a broad class of quantum condensates and ordered systems, providing a  practical computational approach to studying low-temperature ordering dynamics, topological defect evolution, and ultrafast electromagnetic responses of quantum phases in realistic geometries.
\end{abstract}

 \section{Introduction}
\label{sec:level1}

Electromagnetic responses of superconductors underpin a wide range of superconducting electronic devices~\cite{Wang2023, Youssefi2021, McCaughan2019} including quantum technologies and computation~\cite{PhysRevLett.130.047001, Clarke2008, deLeon2021Materials}.  Typically, the low-temperature regime, where the superconducting condensate is fully developed (the order parameter becomes stiff)~\cite{bardeen1957theory,tinkham2004introduction}, is the most technologically relevant temperature regime~\cite{RevModPhys.46.587}.  However, realistic device-level simulations of superconducting ordering dynamics, including vortex formation and their temporal evolution~\cite{Heyl2022,PhysRevB.106.094512,Cordoba-Camacho_2025,3nnt-b8xv}, which determine energy dissipation and device performance in flux-based superconducting circuits~\cite{PhysRevA.107.053704, RevModPhys.85.623, PhysRevLett.128.010604} and quantum technologies~\cite{Houck2012, RevModPhys.86.153}, mostly rely on the Ginzburg-Landau(GL)-formalism-based~\cite{ginzbur1950theory} macroscopic phase-field method~\cite{PhysRevB.44.6916,PhysRevB.47.8016,PhysRevB.52.R15719}. {Correspondingly, the time-dependent Ginzburg-Landau (TDGL) theory~\cite{Schmid1966,BishopVanHorn2023,PhysRevB.46.8376,yang2024optical} extends the GL formalism to real-time dynamics and has been widely used for phase-field simulations of vortex dynamics, phase slips, phase ordering, and nonequilibrium superconducting phenomena, thereby providing the dynamical foundation for modern superconducting phase-field modeling.} The GL theoretical descriptions are, strictly speaking, valid only near the critical temperature $T_c$ and cannot accurately describe the superconducting ordering dynamics in the most operationally relevant low-temperature regime~\cite{bardeen1957theory,tinkham2004introduction,abrikosov2012methods}.  

 Fully microscopic kinetic 
 approaches, such as the Gor'kov Green-function theory of superconductivity~\cite{abrikosov2012methods}, in principle,  provide a rigorous description of electromagnetic responses of superconductors across the entire temperature range. For example,
  a quasiclassical form of this theory, the Eilenberger equation~\cite{eilenberger1968transformation,PhysRevB.95.235403}, can recast the superconducting transport in a more tractable form~\cite{RevModPhys.77.935}, and its diffusive-limit counterpart, the Usadel equation~\cite{PhysRevLett.25.507}, has been widely employed to describe proximity structures and diffusive superconductors~\cite{RevModPhys.77.935}. Beyond these quasiclassical descriptions~\cite{RevModPhys.58.323}, microscopic gauge-invariant kinetic equations~\cite{yang2018gauge,yang2019gauge,PhysRevB.102.014511,PhysRevB.102.144508} have been developed to describe superconducting dynamics directly in terms of gauge-covariant density matrices and kinetic variables, thereby capturing the coupling among quasiparticles, condensate dynamics, and electromagnetic fields in a more transparent manner than conventional pseudospin models~\cite{PhysRevB.91.214505,PhysRevLett.93.160401,PhysRevLett.96.230403,PhysRevB.92.064508}. Nevertheless, the practical implementation of these microscopic kinetic equations for realistic device-level spatiotemporal  simulations is challenging, as it would involve a large number of coupled electronic degrees of freedom and becomes computationally prohibitive.

With the advanced development of quantum technologies over the past few years,  superconducting electronic devices,   including superconducting qubits~\cite{Kjaergaard2020SuperconductingQubits, PhysRevLett.112.190504, Krantz2019QuantumEngineersGuide, Siddiqi2021EngineeringHighCoherence}, hybrid devices~\cite{Frolov2020TopologicalSuperconductivity, DeFranceschi2010Hybrid, Baek2014}, and fluxonic architectures~\cite{Fomin2022CurvedNanoarchitectures, VlaskoVlasov2022TunableMagneticLabyrinth, Golod2015}, operate in the low-temperature regime. Recent  experimental advances~\cite{PhysRevLett.109.187002,ScienceMatsunaga,matsunaga2017polarization,PhysRevLett.122.257001} have  also made it possible to employ intense terahertz (THz) fields to probe and control superconducting ordering dynamics, including manipulating/probing vortex dynamics~\cite{PhysRevLett.133.036004}.  These capabilities offer  prospects for ultrafast switching~\cite{Mitrano2016,Wang2023nc,Budden2021}, coherent control~\cite{Luo2023}, and manipulation of superconducting phases~\cite{Vaswani2021}, with direct implications for ultrafast optics~\cite{PhysRevLett.103.097402, Meng2024, Dong2023Recent, Xu2025Collapse}, THz photonics~\cite{sun2020collective, Welp2013, Kawayama2013THzEmission, Kim2024THzThirdHarmonic}, and quantum-information processing~\cite{PhysRevX.4.031022, PhysRevLett.85.5647, Ourjoumtsev2006SchrodingerKittens, Blais2020CircuitQED}, all of which operate in the low-temperature regime. In this situation, an efficient method capable of bridging microscopic superconducting theories and device-scale spatiotemporal simulations over the entire temperature range from zero temperature to $T_c$ becomes highly desirable.

Motivated by this need, starting from the microscopic BCS mean-field model~\cite{bardeen1957theory,abrikosov2012methods}, we derive a compact phase-field-like dynamical equation for the superconducting order parameter by retaining key microscopic ingredients,  which we refer to as the quantum phase-field formulation. Unlike the traditional phenomenological GL formulation, the quantum phase-field formulation remains applicable well below $T_c$, where superconducting stiffness and phase-coherent dynamics are most relevant. We apply it to simulating the realistic spatiotemporal ordering dynamics under electromagnetic fields and ultrafast THz excitation. It enables real-time simulations of magnetic-field-driven vortex nucleation and motion and captures the ultrafast gap oscillations driven by strong THz fields, all within a unified microscopic quantum description. These features make the quantum phase-field formulation  well suited for modeling superconducting ordering dynamics in realistic device configurations, including vortex control and ultrafast electromagnetic functionality, within a microscopically grounded and internally consistent dynamical framework without relying on phenomenological assumptions.

\section{Methods}\label{sec:level2}

\subsection{Free-energy density functional from microscopic quantum model}

A phase-field model employs an explicit free-energy density functional of a spatially varying complex superconducting order parameter. 
To derive such a functional beyond the conventional phenomenological GL description, we start from the fully microscopic BCS Hamiltonian for a conventional $s$-wave superconductor minimally coupled to an electromagnetic field~\cite{bardeen1957theory,abrikosov2012methods},
\begin{equation}
H=\int{d{\bf x}}\Big[\sum_{\sigma=\uparrow,\downarrow}c^{\dagger}_{\sigma}({\bf x})\left(
\frac{\left[-i\nabla-e{\bf A}({\bf x})\right]^2}{2m}
-\mu
\right)c_{\sigma}({\bf x})-g_{}c^{\dagger}_{\uparrow}({\bf x})c^{\dagger}_{\downarrow}({\bf x})c_{\downarrow}({\bf x})c_{\uparrow}({\bf x})\Big]. 
\end{equation}
Here, $c_{\sigma}({\bf x})$ and $c_{\sigma}^{\dagger}({\bf x})$ denote fermionic annihilation and creation operators for an electron with spin $\sigma$ at position ${\bf x}$. 
The electromagnetic field is introduced via a minimal coupling through the vector potential ${\bf A}({\bf x})$, while $m$ and $\mu$ denote the effective electron mass and chemical potential, respectively. {The coupling constant $g$ represents the effective attractive interaction between electrons in the spin-singlet channel. Within the conventional BCS framework~\cite{bardeen1957theory,abrikosov2012methods}, this interaction originates from phonon-mediated electron--electron attraction. Physically, when two electrons exchange a phonon, an effective attractive interaction is generated between them. This attraction favors the pairing of electrons with opposite momenta and spins, leading to the formation of Cooper pairs. The resulting effective attraction can overcome the Coulomb repulsion at low energies and gives rise to the superconducting state. Therefore, the parameter $g$ characterizes the strength of the pairing interaction responsible for superconductivity.} Throughout this work, we set $\hbar=1$.

Applying the mean-field (Hartree--Fock--Bogoliubov) decoupling to the pairing interaction, one can introduce a spatially dependent superconducting order parameter
\begin{equation}
\Delta({\bf x})
= g\,\langle c_{\downarrow}({\bf x})\,c_{\uparrow}({\bf x}) \rangle ,
\label{eq:gap_def}
\end{equation}
which describes the local Cooper pairing in the $s$-wave spin-singlet channel. Then, the interacting Hamiltonian is reduced to:
\begin{eqnarray}
H_{\rm MF}
&=&
\int d{\bf x}\Bigg[
\sum_{\sigma=\uparrow,\downarrow}
c_{\sigma}^{\dagger}({\bf x})
\left(
\frac{\left[-i\nabla-e{\bf A}({\bf x})\right]^2}{2m}
-\mu
\right)
c_{\sigma}({\bf x})
\nonumber\\
&&\qquad
-\Delta({\bf x})\,
c_{\uparrow}^{\dagger}({\bf x})c_{\downarrow}^{\dagger}({\bf x})
-\Delta^{*}({\bf x})\,
c_{\downarrow}({\bf x})c_{\uparrow}({\bf x})
\Bigg]
+\int d{\bf x}\,\frac{|\Delta({\bf x})|^{2}}{g}.
\label{eq:H_MF_real}
\end{eqnarray}
Writing the complex superconducting order parameter in the amplitude--phase form,
\begin{equation}
\Delta({\bf x}) = |\Delta({\bf x})|\,e^{i\theta({\bf x})},
\end{equation}
we perform a local gauge (unitary) transformation to absorb the phase into the fermionic fields,
\begin{equation}
c_{\sigma}({\bf x}) \rightarrow e^{i\theta({\bf x})/2}\,c_{\sigma}({\bf x}).
\end{equation}
Under this transformation, the pairing term becomes real, while the kinetic term acquires a gauge-invariant superfluid momentum~\cite{yang2018gauge,yang2019gauge,nambu1960quasi,nambu2009nobel},
\begin{equation}
{\bf p}_s({\bf x})=\nabla\theta({\bf x})/2-e{\bf A}({\bf x}).
\end{equation}
Typically, both the amplitude $|\Delta|$ and the phase gradient vary slowly on the scale of the superconducting coherence length.  {More specifically, the characteristic wave vector associated with the spatial variation of the superconducting order parameter satisfies $q \ll k_F$, or equivalently, the variation length scale is much larger than the Fermi wavelength. This condition is generally satisfied in metallic superconductors with a large Fermi surface.} This allows the system to be treated locally as a uniform superconductor within a semiclassical (local equilibrium) approximation. 
In this locally homogeneous limit, momentum remains a good quantum number, and the corresponding mean-field Hamiltonian on this local scale, in momentum space, takes the form
\begin{equation} 
\mathcal{H}={{\sum_{{\bf k}\sigma}}}\xi_{\bf k+p_s}c^{\dagger}_{{\bf k}\sigma}c_{{\bf k}\sigma}-|\Delta|{\sum_{\bf k}}(c_{{\bf k}\uparrow}^{\dagger}c_{-{\bf k}\downarrow}^{\dagger}+c_{-{\bf k}\downarrow}c_{{\bf k}\uparrow})+\frac{|\Delta|^2}{g}, \end{equation} 
where $\xi_{\bf k}={\bf k}^{2}/(2m)-\mu$. Consequently, performing a standard Bogoliubov transformation to diagonalize the mean-field Hamiltonian~\cite{bardeen1957theory,abrikosov2012methods}, we introduce quasiparticle operators
\begin{equation}
\alpha_{\bf k}=u_{\bf k}\,c_{{\bf k}\uparrow}-v_{\bf k}\,c^\dagger_{-{\bf k}\downarrow},\qquad
\beta_{\bf k}=u_{\bf k}\,c_{-{\bf k}\downarrow}+v_{\bf k}\,c^\dagger_{{\bf k}\uparrow},
\end{equation}
with the inverse relations
\begin{equation}
c_{{\bf k}\uparrow}=u_{\bf k}\,\alpha_{\bf k}+v_{\bf k}\,\beta_{\bf k}^\dagger,\qquad
c_{-{\bf k}\downarrow}=u_{\bf k}\,\beta_{\bf k}-v_{\bf k}\,\alpha_{\bf k}^\dagger .
\end{equation}
Here,
\begin{equation}
u_{\bf k}^2=\frac12\left(1+\frac{\xi_{\bf k}}{E_{\bf k}}\right),
\qquad
v_{\bf k}^2=\frac12\left(1-\frac{\xi_{\bf k}}{E_{\bf k}}\right),
\qquad
u_{\bf k}v_{\bf k}=\frac{|\Delta|}{2E_{\bf k}},\qquad E_{\bf k}=\sqrt{\xi_{\bf k}^2+|\Delta|^2}.
\end{equation}
The Hamiltonian is thereby diagonalized as
\begin{equation}
\mathcal{H}_d
=
\sum_{\bf k}
\left(
E_{\bf k}^{+}\,\alpha_{\bf k}^{\dagger}\alpha_{\bf k}
+
E_{\bf k}^{-}\,\beta_{\bf k}^{\dagger}\beta_{\bf k}
\right)-\sum_{\bf k}E_{\bf k}2v_{\bf k}^2+
\frac{|\Delta|^2}{g}+
\frac{n\,p_s^2}{2m},
\label{eq:H_diag}
\end{equation}
where the last term in Eq.~\eqref{eq:H_diag} originates from the diamagnetic contribution in the expansion of $\xi_{{\bf k}+{\bf p}_s}$ with $n$ being the electron density, while the paramagnetic response arises from the Doppler-shifted quasiparticle spectrum, 
\begin{equation}
E_{\bf k}^{\pm}
=
{\bf v}_{\bf k}\!\cdot{\bf p}_s
\pm
E_{\bf k},
\qquad
{\bf v}_{\bf k}=\nabla_{\bf k}\xi_{\bf k}.
\end{equation}

With the diagonalized Hamiltonian, one can proceed with the application of the quantum statistical theory.  Taking into account the presence of a slowly varying phase (superflow), {and evaluating the entropy contribution using the Fermi distribution of  quasiparticles~\cite{abrikosov2012methods}}, the free-energy density functional of the superconducting electrons is obtained as
\begin{eqnarray}
\mathcal{F}^s&=&\langle{\mathcal{H}_d}\rangle-TS=\langle{\mathcal{H}_d}\rangle+k_BT\sum_{{\bf k},\lambda=\pm,\eta=\pm}\big[f({E_{\bf k}^{\lambda}})\ln{f({E_{\bf k}^{\lambda}})+\big(1-f({E_{\bf k}^{\lambda}})\big)\ln\big(1-f({E_{\bf k}^{\lambda}})\big)}\big]\nonumber\\
 &=&-\frac{1}{\beta}\sum_{\mathbf{k},\lambda = \pm} \ln\big(1 \!+\! e^{-\beta(\lambda{\bf v}_{\bf k}\cdot{\bf p}_s+E_{\mathbf{k}})} \big) \!-\! \sum_{\mathbf{k}} E_{\bf k}2v_{\bf k}^2 \!+\! \frac{|\Delta|^2}{g}+\frac{np_s^2}{2m}\nonumber\\
 &\approx&-\frac{2}{\beta}\sum_{\mathbf{k}}\ln\big(1 \!+\! e^{-\beta{E_{\mathbf{k}}}} \big) \!-\! \sum_{\mathbf{k}} \left(E_{\bf k}\!-\! \xi_{\mathbf{k}}\right) \!+\! \frac{|\Delta|^2}{g}+\frac{np_s^2}{2m}+\sum_{\bf k}({\bf v}_{\bf k}\cdot{\bf p}_s)^2\partial_{E_{\bf k}}f(E_{\bf k})\nonumber\\
&=&-\frac{2}{\beta}\sum_{\mathbf{k}}\ln\big(1 \!+\! e^{-\beta{E_{\mathbf{k}}}} \big) \!-\!\sum_{\mathbf{k}} \left(E_{\bf k}\!-\! \xi_{\mathbf{k}}\right) \!+\! \frac{|\Delta|^2}{g}+\frac{n_sp_s^2}{2m}.\label{FEF}
\end{eqnarray}
Here, $f(x)=\frac{1}{e^{\beta{x}}+1}$ is the Fermi distribution; $\beta=1/(k_BT)$ with $k_B$ denoting the Boltzmann constant and $T$ the temperature; $n_s$ represents the superfluid density,
\begin{eqnarray}
n_s&=&n+\frac{2mv_F^2}{3}\sum_{\bf k}\partial_{E_{\bf k}}f(E_{\bf k})=\frac{2k_F^2}{3m}\sum_{\bf k}\Big[-\partial_{\xi_{\bf k}}\Big(\frac{\xi_{\bf k}}{E_{\bf k}}\frac{2f(E_{\bf k})-1}{2}\Big)+\partial_{E_{\bf k}}f(E_{\bf k})\Big]\nonumber\\
&=&\frac{2k_F^2}{3m}\sum_{\bf k}\Big[-\frac{|\Delta|^2}{E^3_{\bf k}}\frac{2f(E_{\bf k})-1}{2}-\frac{\xi^2_{\bf k}}{E^2_{\bf k}}\partial_{E_{\bf k}}f(E_{\bf k})+\partial_{E_{\bf k}}f(E_{\bf k})\Big]
\nonumber\\
&=&|\Delta|^2\frac{2k_F^2}{3m}\sum_{\bf k}\frac{\partial_{E_{\bf k}}}{E_{\bf k}}\Big[\frac{2f(E_{\bf k})-1}{2E_{\bf k}}\Big].
\end{eqnarray}
{Apart from the superflow contribution ${n_sp_s^2}/({2m})$, the superconducting-electron free energy (i.e., the electronic contribution to the free energy) derived here has the same form  as the corresponding electronic contribution obtained in Ref.~\cite{PhysRevB.58.9365}. The total free energy of the superconducting system additionally includes the standard electromagnetic-field energy term, which is not explicitly written in Eq.~(\ref{FEF}), since its variation with respect to the vector potential simply gives rise to the standard Maxwell equation, which is solved directly in the present work.}  The expression of the superfluid density $n_s$ here is also identical to the standard microscopic expressions obtained  in various microscopic approaches (such as Green-function methods~\cite{abrikosov2012methods}, kinetic approaches~\cite{yang2018gauge,yang2019gauge} as well as path-integral formalism~\cite{yang2024diamagnetic,sun2020collective}).

We write $\Delta({\bf x})=|\Delta|e^{i\theta({\bf x})}$, 
the free-energy functional in Eq.~(\ref{FEF}) can be recast in the form
\begin{eqnarray}
\mathcal{F}^s&=&-\frac{2}{\beta}\sum_{\mathbf{k}}\ln\big(1 \!+\! e^{-\beta{E_{\mathbf{k}}}} \big) \!-\! \sum_{\mathbf{k}} \left(E_{\bf k}\!-\! \xi_{\mathbf{k}}\right) \!+\! \frac{|\Delta|^2}{g}+\frac{\lambda_s|(\nabla-2ie{\bf A})\Delta|^2}{4m}, \label{FE}
\end{eqnarray}
with
\begin{equation}
\lambda_s=\frac{k_F^2}{3m}\sum_{\bf k}\frac{\partial_{E_{\bf k}}}{E_{\bf k}}\Big[\frac{2f(E_{\bf k})-1}{2E_{\bf k}}\Big].
\end{equation} 
The last gradient term represents the phase stiffness and describes the energy cost associated with spatial variations of the superconducting order parameter.
In the phase-dominated regime, where the amplitude $|\Delta|$ is nearly constant on local length scales, this term reduces to a London-type form proportional to $(\nabla\theta-2e{\bf A})^2$, thereby directly recovering the Meissner screening. This form follows directly from the gauge-invariant coupling between the superconducting order parameter and the electromagnetic field~\cite{abrikosov2012methods}. Consequently, the microscopic BCS theory naturally yields the standard low-energy hydrodynamics of superconductivity~\cite{abrikosov2012methods}.
This free-energy functional therefore provides a microscopic starting point for the quantum phase-field formulation.

\subsection{Dynamic quantum phase-field model}
\label{secMiPF}

Given the derived free-energy functional, the spatiotemporal evolution of the superconducting order parameter $\Delta$ can be described within a coarse-grained effective field theory.
In this framework, the dynamics of $\Delta$ is governed by a generalized  Euler--Lagrange equation for a quantum field with both inertial and dissipative terms~\cite{peskin2018introduction}
\begin{equation}
  u\partial_t^2\Delta+\gamma\partial_t\Delta\!=-\frac{\partial{\mathcal{F}^s}}{\partial{\Delta}^*}+D_{\mu}\Big(\frac{\partial{\mathcal{F}^s}}{\partial(D_{\mu}\Delta^*)}\Big), \label{eq:ELeq}
\end{equation}
where $u$ and $\gamma$ are phenomenological parameters characterizing the inertial and dissipative responses of the order-parameter field, and the covariant derivative is defined as $D_\mu=\nabla-2ie{\bf A}$. Substituting the explicit form of $\mathcal{F}^s$ into Eq.~\eqref{eq:ELeq}, one obtains the quantum phase-field equation 
\begin{equation}\label{GapEq}
u\partial_t^2\Delta+\gamma\partial_t\Delta=\frac{\lambda_s(\nabla-2ie{\bf A})^2\Delta}{4m}-\frac{\Delta}{g}+\sum_{\bf k}\frac{\Delta[1-2f(E_{\bf k})]}{2E_{\bf k}},
\end{equation}
which generalizes the time-dependent Ginzburg--Landau equation by incorporating the full microscopic BCS gap kernel. For clarity and practical use, we again explicitly give the expressions for $\lambda_s$ and $E_{\bf k}$ below:
\begin{equation}
E_{\bf k}=\sqrt{\xi_{\bf k}^2+|\Delta|^2},
\qquad
\lambda_s=\frac{k_F^2}{3m}\sum_{\bf k}\frac{\partial_{E_{\bf k}}}{E_{\bf k}}\Big[\frac{2f(E_{\bf k})-1}{2E_{\bf k}}\Big].
\end{equation}

Physically, the dynamical equation~(\ref{GapEq}) describes the spatiotemporal evolution of the complex superconducting condensate as a coarse-grained quantum field, incorporating both inertial and dissipative dynamics~\cite{pekker2015amplitude}. 
The second-order time derivative encodes the reversible, propagating response of the condensate, while the linear damping term captures energy relaxation arising from coupling to fermionic quasiparticles and other microscopic degrees of freedom.
The gauge-covariant gradient structure ensures that the condensate dynamics is consistently coupled to the electromagnetic field, reflecting the underlying local $U(1)$ symmetry of a charged complex order parameter. Importantly, the fermionic contribution entering through the momentum summation retains the explicit information about the BCS quasiparticle spectrum, thereby anchoring the effective phase-field dynamics to its microscopic origin. As a result, the present formulation goes beyond purely phenomenological approaches by preserving the gauge consistency and microscopic input for both statics and dynamics.

In contrast to the conventional time-dependent Ginzburg--Landau theories, which are formulated in the vicinity of the superconducting transition temperature $T_c$, the quantum phase-field framework remains applicable deep in the superconducting state.
This enables direct real-space and real-time simulations of order-parameter dynamics far below $T_c$, where phase stiffness is large and collective phenomena such as vortex motion, amplitude (Higgs) modes, and ultrafast electromagnetic responses dominate the superconducting behavior.

\section{Results}
\label{sec:level3}

Having established the equation of motion for the superconducting order parameter  in Sec.~\ref{secMiPF}, we now turn to numerical simulations of electromagnetic responses over the entire temperature range, from deep in the superconducting phase up to the transition temperature $T_c$.
Our simulations address both static and dynamical nonequilibrium phenomena that are central to superconducting electrodynamics. We consider two representative nonequilibrium electromagnetic scenarios.
First, we investigate the magnetic response of type-II superconductors under static magnetic fields, with particular emphasis on vortex nucleation, spatial ordering, and real-time vortex dynamics.
Second, we study the ultrafast optical response driven by intense THz fields, where the superconducting condensate is driven far from equilibrium and exhibits coherent oscillations associated with collective amplitude (Higgs) dynamics.

In all simulations, the superconducting order parameter $\Delta({\bf x},t)$ evolves according to the gauge-covariant dynamical equation~(\ref{GapEq}), while electromagnetic fields are incorporated through the vector potential ${\bf A}({\bf x},t)$.
The microscopic input encoded in the free-energy functional, including the electronic dispersion and pairing interaction, ensures that both equilibrium properties and nonequilibrium dynamics remain consistent with the underlying BCS quasiparticle spectrum across the entire temperature range.

Without loss of generality, and to facilitate comparison among different superconducting systems (conventional superconductors), all results presented in this work are expressed in normalized units. In this formulation, material-specific parameters, such as the sample size, pairing interaction strength, effective electron mass, and electronic-band parameters, enter only through dimensionless combinations and can be absorbed into the choice of units.  Consequently, variations of these microscopic parameters merely rescale the quantitative values of the observables but do not modify the qualitative physical conclusions drawn in this work.

\subsection{Vortex nucleation and dynamics}

We first investigate the response of the superconducting order parameter to a static magnetic field.  In addition to the dynamical equation of motion for the order parameter, the electromagnetic field evolves self-consistently according to the Maxwell equation for the vector potential~\cite{Heyl2022,PhysRevB.106.094512,abrikosov2012methods}.  {Specifically, starting from the Maxwell equation for the vector potential,
\begin{equation}
\nabla\times\nabla\times{\bf A}=
\mu_0{\bf J}_s,
\end{equation}
the superconducting current density is obtained from the variation of the superconducting-electron free energy with respect to the vector potential,
\begin{equation}
{\bf J}_s=-\frac{\delta F_s}{\delta{\bf A}} .
\end{equation}
Using Eq.~(\ref{FEF}), this gives
\begin{equation}
{\bf J}_s
=-\frac{e\lambda_s}{m}\Big[2e{\bf A}|\Delta|^{2}+\frac{i}{2}
\left(
\Delta^{*}\nabla\Delta
-
\Delta\nabla\Delta^{*}
\right)
\Big],
\end{equation} 
where the second term on the right-hand side represents the superconducting current generated by the phase stiffness of the condensate. 
Substituting this expression into the Maxwell equation yields the standard vector-potential equation,
\begin{equation}
\nabla\times\nabla\times{\bf A}
+\frac{e\mu_0\lambda_s}{m}\Big[2e{\bf A}|\Delta|^{2}+\frac{i}{2}
\left(
\Delta^{*}\nabla\Delta
-
\Delta\nabla\Delta^{*}
\right)
\Big]
=
0.
\end{equation} 
For numerical implementation in the phase-field simulation, the stationary Maxwell equation is solved using a relaxation dynamics,} 
\begin{equation}
\partial_t{\bf A}
=\nabla\times\nabla\times{\bf A}
+\frac{e\mu_0\lambda_s}{m}\Big[2e{\bf A}|\Delta|^{2}+\frac{i}{2}
\left(
\Delta^{*}\nabla\Delta
-
\Delta\nabla\Delta^{*}
\right)
\Big],\label{eqMax}
\end{equation}
 whose steady-state solution recovers the original Maxwell equation. The simulations are performed with physically transparent boundary conditions~\cite{Heyl2022,PhysRevB.106.094512,Cordoba-Camacho_2025,3nnt-b8xv}. 
At the sample boundary, the superconducting order parameter satisfies the Neumann-type condition~\cite{PhysRevB.44.6916,PhysRevB.46.8376,PhysRevB.47.8016,PhysRevB.52.R15719}
\begin{equation}
(\nabla - 2 i e {\bf A}) \Delta \cdot {\bf n} = 0,\label{eqb1}
\end{equation}
which prohibits supercurrent flow normal to the surface. 
Meanwhile, the magnetic boundary condition is imposed as
\begin{equation}
(\nabla \times {\bf A}) \times {\bf n} = {\bf B} \times {\bf n},\label{eqb2}
\end{equation}
ensuring that the externally applied magnetic field ${\bf B}$ penetrates the system consistently through the vector potential. 

To probe vortex nucleation and magnetic relaxation, we initialize the system with a spatially uniform superconducting order parameter,
$\Delta({\bf x},t=0)=\Delta_0$,
together with a vector potential corresponding to the externally applied magnetic field ${\bf B}$.
The system is then allowed to evolve self-consistently according to the coupled order-parameter equation of motion and Maxwell equations until convergence is reached. Importantly, no artificial random perturbations are introduced in the initial state.
Instead, vortex nucleation and subsequent magnetic relaxation emerge dynamically from the intrinsic instability of the uniform condensate under the applied field.
For this purely relaxational condensate dynamics, we set the inertial coefficient $u=0$ and choose a finite dissipative parameter $\gamma$, which ensures numerical stability while driving the system toward equilibrium. This procedure enables the condensate to relax naturally into its equilibrium or metastable magnetic configuration, allowing vortex structures to form spontaneously without imposing any \emph{a priori} assumptions about their number, position, or geometry.

\begin{figure}[h]
\centering
\includegraphics[width=15.6cm]{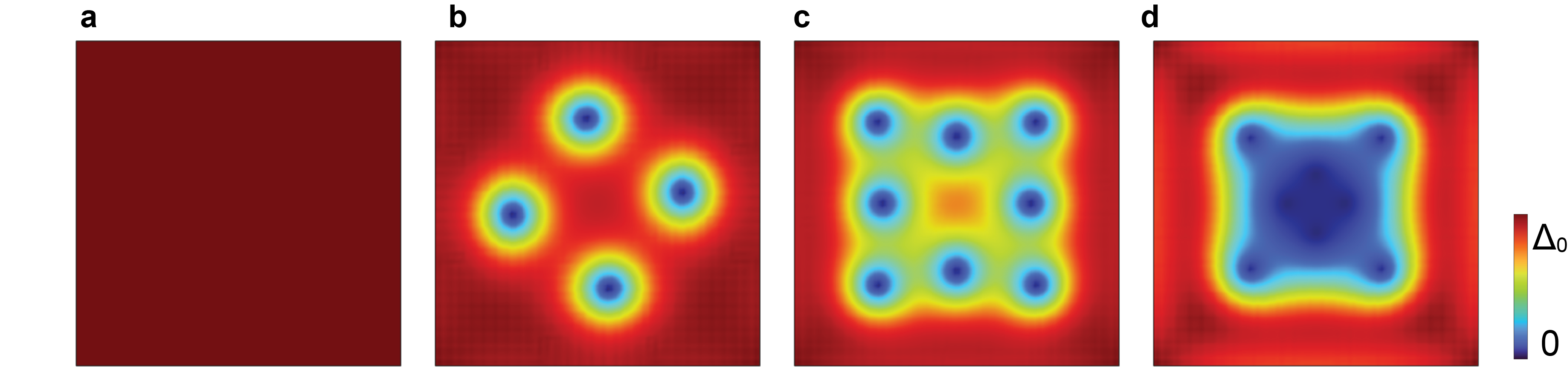}
\caption{Steady-state spatial distributions of the superconducting order-parameter amplitude
$|\Delta(\mathbf{x})|$ at high temperature $T/T_c=0.7$ under applied magnetic fields
$\bar{B}/H_{c2}(0)=0.09$, $0.22$, $0.40$, and $0.51$. $\Delta_0=|\Delta(T=0)|$ and $H_{c2}(0)=H_{c2}(T=0)$.}
\label{fig:high-temp}
\end{figure}

Figure~\ref{fig:high-temp} shows the numerical results  of steady-state spatial distributions of the superconducting
order-parameter amplitude $|\Delta(\mathbf{x})|$ at a high reduced temperature $T/T_c=0.7$
for several values of the applied magnetic field. At this temperature, which is sufficiently
close to the superconducting transition temperature $T_c$, the system is expected to be well described by the conventional GL formulation.
 Consistent with this expectation, the  quantum phase-field simulations reproduce the standard sequence
of magnetic-field-driven states of a type-II superconductor. At low fields
[$\bar{B}/H_{c2}(0)=0.09$], the system remains in a nearly homogeneous superconducting state.
With increasing magnetic field [$\bar{B}/H_{c2}(0)=0.22$], isolated vortices nucleate and form
a dilute vortex state, characterized by well-separated vortex cores. Upon further increasing the magnetic field [$\bar{B}/H_{c2}(0)=0.40$], vortex–vortex interactions
and thermal fluctuations lead to strong positional disorder, resulting in a vortex-liquid-like
state in which vortices lose long-range positional order while maintaining a finite local
superconducting amplitude. Finally, at sufficiently large fields
[$\bar{B}/H_{c2}(0)=0.51$], superconductivity is strongly suppressed and the system crosses over
to a normal state with vanishing $|\Delta(\mathbf{x})|$ in the most areas. These results demonstrate that, in the high-temperature regime close to $T_c$, the quantum
phase-field model smoothly connects to the classical GL description.

\begin{figure}
\centering
\includegraphics[width=15.7cm]{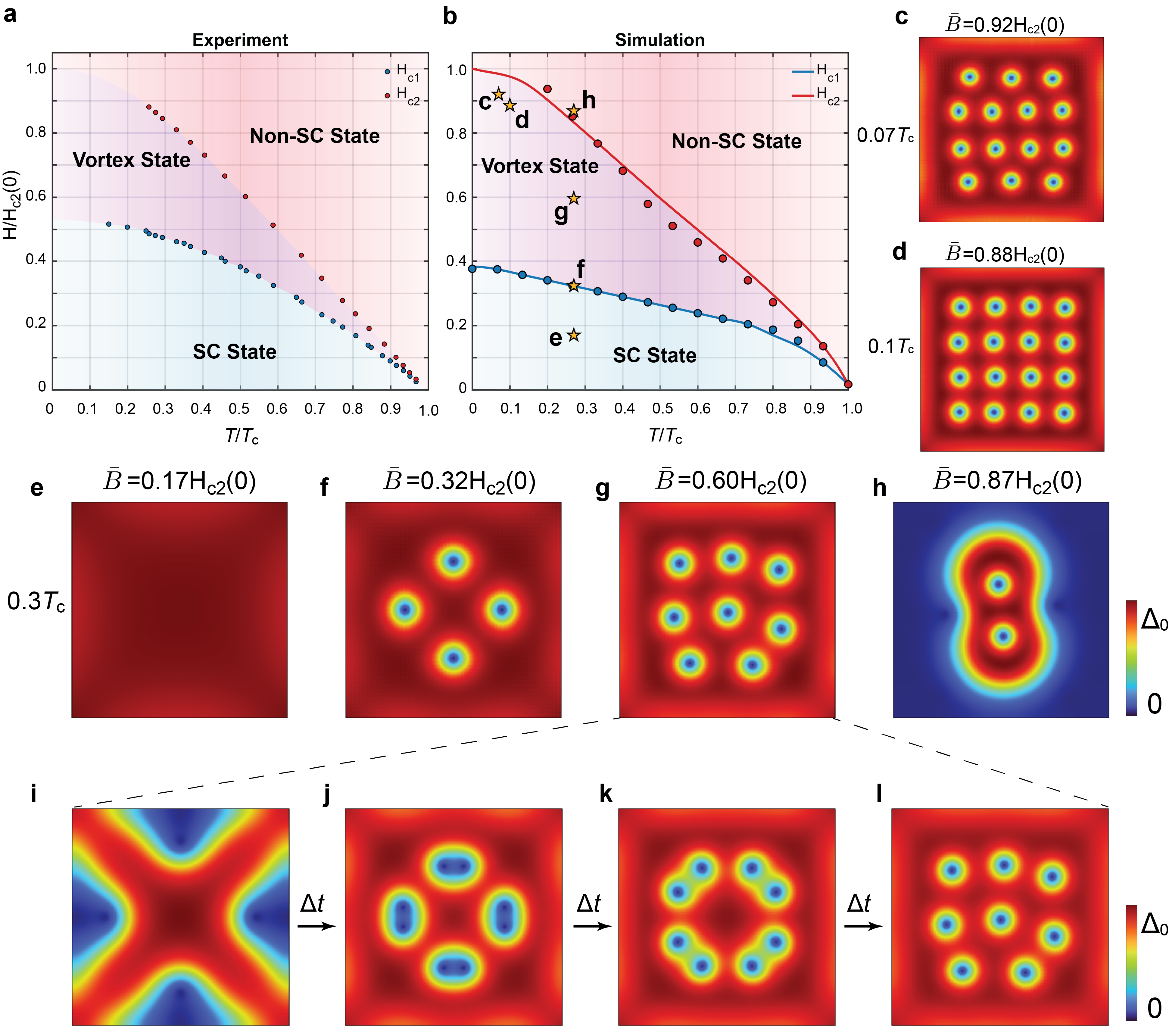}
\caption{Magnetic response and vortex dynamics of a type-II superconductor obtained from the quantum phase-field simulations. 
Temperature dependence of the lower and upper critical magnetic fields $H_{c1}$ and $H_{c2}$ extracted from {\bf a} experimental measurements of Ref.~\cite{Sekula}, and {\bf b} numerical simulations based on the quantum phase-field model. Dots in {\bf b} denote simulation data, while the solid curves serve as guides to the eye. {\bf c} and {\bf d} demonstrate steady-state spatial distributions of the superconducting order-parameter amplitude $|\Delta({\bf x})|$ at low temperatures $T/T_c=0.07$ and $T/T_c=0.1$ under magnetic fields inside the vortex state.
{\bf e}-{\bf h} display $|\Delta({\bf x})|$ at $T/T_c=0.3$ under increasing magnetic-field strengths. 
{\bf i}-{\bf l} show real-time evolution of spontaneous vortex formation corresponding to the steady-state configuration shown in panel {\bf g}. Simulation parameters and numerical treatment are described in the Appendix.} \label{fig:pinning}
\end{figure}

The results of the numerical simulations of full temperature range are summarized in Fig.~\ref{fig:pinning}.
As shown in Fig.~\ref{fig:pinning}(b), the lower and upper critical fields, $H_{c1}$ and $H_{c2}$, extracted from our simulations exhibit temperature dependencies that are in good agreement with experimental measurements of Ref.~\cite{Sekula}, reproduced in Fig.~\ref{fig:pinning}(a).
 This level of agreement is achieved within the quantum phase-field framework using only microscopic input parameters, without resorting to the explicit real-space time evolution of microscopic density matrices. Moreover, as shown in Fig.~\ref{fig:pinning}(c)~and~(d), the method can yield vortex nucleation and spatial organization at low temperatures far below $T_c$, showing its applicability over a wide temperature range from 0 to $T_c$.

To illustrate the magnetic-field-driven evolution of the superconducting state at low temperatures, we present the steady-state order-parameter configurations at $T/T_c=0.3$ under different magnetic-field strengths in Figs.~\ref{fig:pinning}(e)--(h).  These results demonstrate that vortex nucleation and spatial organization emerge self-consistently from the competition between magnetic screening and superconducting phase stiffness. At low magnetic fields [Fig.~\ref{fig:pinning}(e)], the system remains in a homogeneous superconducting state, characterized by a nearly uniform gap amplitude and complete magnetic screening. 
Upon increasing the magnetic field [Fig.~\ref{fig:pinning}(f)], vortices nucleate and the system enters the mixed (vortex) state, where superconductivity persists in the background while quantized flux lines penetrate the sample.  Figs.~\ref{fig:pinning}(i)--(l) illustrate the dynamical process of spontaneous vortex formation~\cite{PhysRevB.44.6916,PhysRevB.46.8376,PhysRevB.47.8016,PhysRevB.52.R15719} when the local magnetic field strength exceeds the stability threshold. Starting from a homogeneous superconducting state, the magnetic field initially penetrates the sample from the boundaries, where superconducting screening currents are weakest. When magnetic flux penetrating from different boundaries intersects, topological phase singularities nucleate and subsequently evolve into quantized vortices. 
These vortices then propagate and spread into the bulk, driven by the competition between magnetic flux penetration and superconducting phase stiffness, eventually organizing into a stable vortex configuration. Importantly, this vortex formation pathway, triggered by the intersection of magnetic flux penetrating from different boundaries, is not a numerical artifact but a direct consequence of the topological constraint imposed by the single-valued superconducting order parameter. 
This is because the phase of the complex order parameter must wind by integer multiples of $2\pi$, making the nucleation of vortices unavoidable once the local stability threshold (local singularity) is exceeded. 
The quantum phase-field framework captures this process naturally through the continuous evolution of the complex order parameter, without manually introducing the vortices.

Remarkably, at even higher magnetic fields [Figs.~\ref{fig:pinning}(g) and \ref{fig:pinning}(h)], superconductivity is progressively suppressed, and the system evolves toward a predominantly normal-state background, within which isolated superconducting regions, and a small number of residual vortices remain. This high-field regime reflects the gradual breakdown of global phase coherence rather than an abrupt disappearance of local pairing. 
While the magnetic field suppresses the long-range superconducting order by destroying phase stiffness, remnants of the pairing amplitude survive locally, giving rise to isolated superconducting regions embedded in a predominantly normal-state background. 
This behavior of mixed state consisting of superconducting island 
 is consistent with the previous microscopic calculation~\cite{dubi2007nature} in which the superconducting transition under strong magnetic fields is driven by phase disordering and superconducting islands are surrounded by normal-state regions. Notably, this inhomogeneous, island-dominated regime is a characteristic feature of the low-temperature phase.  In contrast, at higher temperatures (see Fig.~\ref{fig:high-temp}),  thermal proliferation and merging of vortices strongly reduce the effective local superconducting gap in their vicinity through vortex-core overlap and enhanced amplitude fluctuations, thereby suppressing global superconductivity and giving rise to a vortex-liquid–like state, in which the physics is vortex-dominated and superconducting islands no longer form. Therefore, the emerging low-temperature mixed state represents a reversal of the conventional
host–defect structure of the vortex state: superconductivity survives in the
form of phase-coherent islands embedded in a normal background, with vortices
nucleating only within these islands. {In general, the standard homogeneous GL framework is not the most natural description of such an island-dominated mixed state, while similar phenomena, including island-like superconducting structures and localized vortex nucleation, can emerge in extended GL/TDGL formulations~\cite{Schmid1966,BishopVanHorn2023,Wang2017CGA,PhysRevB.107.144514} incorporating spatial inhomogeneity, granularity, defects, or proximity-coupled superconducting regions. Consequently, the standard homogeneous GL framework is a less suitable baseline for describing this regime.}

These results demonstrate the application of quantum phase-field framework as a microscopic yet computationally efficient tool for simulating the magnetic response and vortex formation across the entire superconducting temperature range.

\subsection{THz-optical response: ultrafast gap dynamics}

We next turn to the ultrafast optical response of superconductors driven by intense THz fields. 
In contrast to the quasi-static magnetic response discussed above, THz excitation probes the intrinsic inertial dynamics of the superconducting condensate on sub-picosecond time scales, where the temporal evolution of the gap amplitude plays a central role.  In a typical pump--probe experiment~\cite{shimano2020higgs}, a strong THz pump pulse ${\bf A}_{\rm pump}$ excites the nonequilibrium gap dynamics of the form
\begin{equation}
|\Delta(t)| = |\Delta_0| + \delta|\Delta(t)|,
\end{equation}
which are subsequently detected by a weak probe pulse. 
Within the quantum phase-field framework, the probe signal arises from the modulation of the supercurrent induced by the pump-driven gap dynamics. 
Specifically, the pump-induced changes in the probe electric field satisfy
\begin{equation}
\delta E_{\mathrm{probe}} \propto \delta {\bf j}_s,
\end{equation}
where ${\bf j}_s$ denotes the superconducting current density. 
Using ${\bf j}_s =- \lambda_s |\Delta|^2 e^2 {\bf A}_{\mathrm{probe}}/m$, the pump-induced variation of the supercurrent can be expressed as~\cite{yang2024optical,yang2023optical,cui2019impact,shimano2020higgs}
\begin{equation}
\delta {\bf j}_s
=
-\delta\!\left(\lambda_s |\Delta|^2 e^2 {\bf A}_{\mathrm{probe}}/m\right)
\simeq-
\lambda_s |\Delta_0| \, \delta|\Delta(t)| \, e^2 {\bf A}_{\mathrm{probe}}/m,
\end{equation}
where we have retained the leading-order contribution in the gap-amplitude fluctuation $\delta|\Delta(t)|$.

In this regime, the inertial term in the dynamic equation of motion becomes essential, and therefore we  retain a finite coefficient $u \neq 0$ to capture the ultrafast collective dynamics of the order parameter. 
Specifically, the coefficient $u$ is given microscopically by~\cite{yang2024optical,yang2023optical,yang2019gauge}
\begin{equation}
u=\sum_{\bf k}\frac{\partial_{E_{\bf k}}}{2E_{\bf k}}
\Bigg[\frac{2f(E_{\bf k})-1}{2E_{\bf k}}\Bigg],
\end{equation}
which is directly determined by the underlying BCS quasiparticle spectrum. With this microscopic input, the nonequilibrium gap dynamics under THz excitation can be obtained by self-consistently solving the equation of motion [Eq.~(\ref{GapEq})] in the presence of the time-dependent electromagnetic field ${\bf A}(t)$. The relation between the temporal and spatial gradient coefficients $\lambda_s=u2k_F^2/(3m)$ ensures an emergent Lorentz covariance of the quantum phase-field equation, leading to a relativistic Klein–Gordon–type dynamics of the superconducting order parameter with an effective velocity of $c_{\Delta}=\sqrt{2/3}v_F$.

\begin{figure}
\centering
\includegraphics[width=15.7cm]{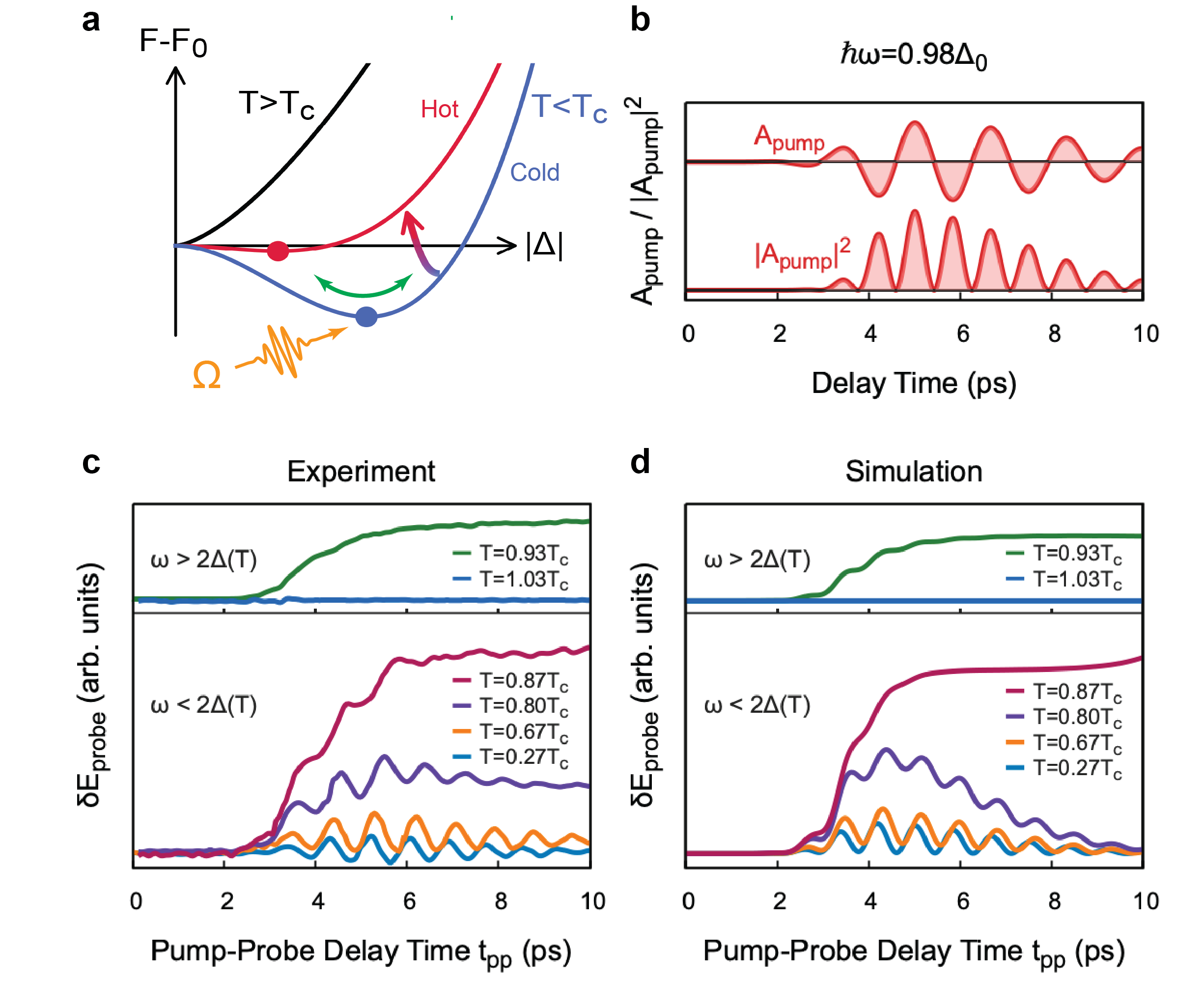}
\caption{
Ultrafast THz-driven gap dynamics of a superconductor probed by time-resolved pump--probe spectroscopy. 
{\bf a} is a schematic illustration of the nonequilibrium free-energy landscape under strong THz excitation, where nonlinear energy injection substantially modifies the superconducting state and the THz field drives collective amplitude dynamics. 
{\bf b} is the temporal waveform of the applied multi-cycle THz pump pulse $A_{\mathrm{pump}}(t)$ with a center frequency of $\omega\sim0.98|\Delta(T=0)|$ in experiment (Ref.~\cite{ScienceMatsunaga}) and simulation; the squared field $|A_{\mathrm{pump}}(t)|^2$  is also shown. 
{\bf c} is the experimentally measured pump--probe response $\delta E_{\mathrm{probe}}(t_{\mathrm{pp}})$ as a function of the pump--probe delay $t_{\mathrm{pp}}$ on NbN thin films. The data come from Ref.~\cite{ScienceMatsunaga}. 
{\bf d} shows the corresponding pump--probe response $\delta E_{\mathrm{probe}}(t_{\mathrm{pp}})$ obtained from the quantum phase-field simulations, capturing the coherent oscillatory gap dynamics induced by the THz pump field. Simulation
parameters and numerical treatment are described in the Appendix. }

\label{fig:THz}
\end{figure}

In the specific simulations, we consider a multi-cycle THz pump pulse ${\bf A}(t)=A_{\rm pump}(t)$, as experimentally applied and shown in Fig.~\ref{fig:THz}{\bf b}. 
Within the quantum phase-field framework, the THz field is coupled to the superconducting condensate through the time-dependent vector potential ${\bf A}_{\rm pump}(t)$ in a fully gauge-invariant manner. 
This coupling naturally drives the superconducting order parameter out of equilibrium and gives rise to coherent oscillations of the gap amplitude $|\Delta(t)|=|\Delta_0|+\delta|\Delta(t)|$, associated with the collective Higgs-like modes of the condensate~\cite{ScienceMatsunaga,PhysRevLett.120.117001,matsunaga2013higgs}. Importantly, the quantum phase-field formulation enables direct real-time simulations of such ultrafast gap dynamics in real space, without resorting to pseudospin representations or solving microscopic kinetic equations.

Specifically, Figs.~\ref{fig:THz}(c) and \ref{fig:THz}(d) present the experimentally measured time-resolved pump--probe response of NbN under THz excitation~\cite{ScienceMatsunaga} and the corresponding theoretically calculated response obtained from the scaled quantum phase-field model at different temperatures, respectively. At low temperatures below $T<0.8T_c$, the superconducting condensate is well developed, and the population of thermally excited Bogoliubov quasiparticles is strongly suppressed. 
In this regime, the THz field can coherently drive the Higgs (amplitude) mode of the superconducting order parameter, giving rise to a predominantly oscillatory response of the superconducting gap. This oscillatory signal closely follows the temporal profile of $A_{\mathrm{pump}}^2(t)$, reflecting the parametric coupling between the THz field and the amplitude mode of the superconducting condensate. 
This behavior is characteristic of a nonlinear coherent excitation of the Higgs mode in a well-established superconducting state.

As the temperature increases above $\sim 0.8T_c$, thermal quasiparticle excitations become increasingly important and modify the nonequilibrium response of the system under a strong THz driving field. 
The coherent Higgs oscillations gradually weaken and develop on top of an emerging non-oscillatory background. 
This background originates from incoherent heating induced by the nonlinear THz excitation, as schematically illustrated in Fig.~\ref{fig:THz}{\bf a}. 
In this regime, the energy injected by the THz field effectively drives the system toward a higher free-energy state, reducing the relative weight of coherent amplitude oscillations while enhancing the non-oscillatory response. Above $T_c$, long-range superconducting order is destroyed and the gap amplitude collapses, eliminating the collective Higgs mode. 
As a result, the superconducting contribution to the pump--probe response vanishes and no oscillatory signal is observed, consistent with the disappearance of the superconducting condensate.

\section{Summary}

 From a theoretical perspective, since the phenomenological GL theory near $T_c$ has played a central role in superconducting engineering over the past decades, a closely analogous but non-phenomenological description, while retaining direct microscopic consistency beyond the immediate vicinity of the critical temperature, is expected to be particularly valuable for device-scale applications.
 It bridges the fundamental, spatially uniform microscopic BCS theory and a macroscopic phase-field formulation of superconducting ordering dynamics, and therefore offers a practical, microscopically grounded modeling tool for superconducting systems across a wide range of temperatures, time scales, and electromagnetic driving conditions.

We have applied this formulation to two representative and technologically relevant scenarios: magnetic-field-driven vortex dynamics and ultrafast THz excitation. For static magnetic fields, the simulations capture vortex nucleation, motion, and spatial organization self-consistently, reproducing the temperature dependence of the lower and upper critical fields in quantitative agreement with experimental measurements. Notably, the vortex dynamics emerge naturally from the phase-field description without invoking phenomenological vortex-core models. For ultrafast THz excitation, the quantum phase-field approach incorporates inertial condensate dynamics and enables direct simulations of time-resolved pump–probe responses associated with collective gap-amplitude (Higgs-like) modes. The calculated responses reproduce key experimental trends, including coherent oscillations at low temperatures, their nonlinear evolution under strong driving at elevated temperatures, and the disappearance of the signal above $T_c$. Together, these results demonstrate the capability of the quantum phase-field formulation to capture superconducting ordering dynamics across disparate spatial and temporal regimes. Relative to fully microscopic dynamical simulations involving high-dimensional electronic density matrices and kinetic variables~\cite{PhysRevB.97.205301,RevModPhys.77.1321,RevModPhys.77.935}, its favorable computational efficiency naturally enables practical extensions to complex geometries and hybrid superconducting structures of direct relevance to superconducting technologies.

\section{Acknowledgments}

This work was supported by the U.S. Department of Energy, Office of Science, Basic Energy Sciences, under Award No. DE-SC0020145, as part of the Computational Materials Sciences Program. F.Y. and L.Q.C. also acknowledge the generous support of the Donald W. Hamer Foundation through a Hamer Professorship at Penn State.

\section{Data availability}
All data that support the findings of this study are included
within the article (and any supplementary files).

\section{Appendices}

\appendix

\section{Computation of the phase diagram}

For the response to a static magnetic field,  we performed a finite element analysis (FEA) by solving Eq.~(\ref{GapEq}) and Eq.~(\ref{eqMax}) subject to the boundary conditions Eqs.~(\ref{eqb1}) and~(\ref{eqb2}). To ensure the uniqueness of the numerical solution for the vector potential $\mathbf{A}$ without affecting the physical magnetic field $\nabla\times\mathbf{A}$, we additionally imposed the gauge condition $\mathbf{A\cdot n}=0$ on the boundary. The problem was implemented and solved using the PDE Module in COMSOL\cite{comsol_multiphysics_64}. Calculations were carried out in Gaussian units with $\bar{L}$=1~\AA, $\bar{E}=1$~eV, and $\bar{T}=1~$ps. The computational domain was a two-dimensional rectangle of 400~nm$\,\times\,$400~nm, discretized using a free quadrilateral mesh with a maximum element size of 4 nm.

All model parameters are summarized in Table~\ref{tab:pf_params}. To reduce the computation cost associated with calculating the momentum summation for $\lambda_s(|\Delta|)$, $F(|\Delta|)=\sum_{\bf k}\frac{1-2f(E_{\bf k})}{2E_{\bf k}}$, we precomputed these quantities on a discrete grid. During simulation, $\lambda_s(\Delta)$ and $F(|\Delta|)$ were evaluated using COMSOL interpolation functions. Without any parallelization, each simulation at a fixed temperature, in which the magnetic-field range was swept, typically required several hours, with lower-temperature cases being more computationally demanding. Moreover, since our primary interest lies in equilibrium properties, we focused on stationary solutions and neglected kinetic evolution when constructing the phase diagram. In addition to low-temperature cases in Fig.~\ref{fig:pinning}, Fig.~\ref{fig:high-temp} displays a higher-temperature case when $T/T_c=0.7$ where the vortex size is larger.

\begin{table}[htbp]
  \centering
  \caption{\small{Parameters used in the numerical simulations.
   In the numerical simulations, the momentum summation is replaced by $\sum_{\bf k}\rightarrow{D\int^{\omega_D}_{-\omega_D}{d\xi_{\bf k}}}$, where $D$ denotes the density of states at the Fermi level and $\omega_D$ is the Debye energy cutoff in the conventional BCS theory. Here, $m_e$ denotes the free-electron mass.  The pairing potential 
   $gD$ and $\omega_D$ are determined by fitting to the experimentally measured superconducting transition temperature and zero-temperature gap of NbN thin films in Ref.~\cite{ScienceMatsunaga},
    $T_c\approx15.2~$K and $|\Delta(T=0)|=2.53~$meV. Moreover, we adopt a large Fermi energy, representative of a  metallic system. Within this parameter regime, the resulting zero-temperature upper critical field is $H_{c2}(T=0)=0.31$-0.32~T. Without loss of generality, and to facilitate comparison among different superconducting systems (conventional superconductors), all results presented in this work are expressed in normalized units. Under the normalization convention here, the quantum phase-field model in Eq.~(\ref{GapEq}) is unaffected by the specific value of the normal-state density of states.}}
  \label{tab:pf_params}
  \begin{tabular}{lllll}
    \hline
    \hline
    $m$ & \quad 
    $\omega_D$ &  \quad 
    $gD$ &\quad  $Dk_F^2/(3m)$ \\
$m_e$ &\quad   $2.6~\mathrm{meV}$ & \quad   $1.11$  
   &\quad  $1.6\times10^{14}/$cm$^2$ \\
    \hline
    \hline
  \end{tabular}
\end{table}

\section{Computation of ultrafast optical responses}

When we calculated the time-resolved pump–probe responses under THz excitation, we assumed the material to be homogeneous. From the experiment \cite{ScienceMatsunaga}, we fitted 
\begin{equation}
A_{\mathrm{pump}}(t)=A_0 \exp(-(t-t_0)^2/\sigma^2) \cos(\omega t+\phi),
\end{equation}
with $\sigma=\sigma_1$ at $t<t_0$, $\sigma=\sigma_2$ at $t>t_0$, where $\omega=2\pi f$ ($f=0.6~$THz), $\sigma_1=1.341~$ps, $\sigma_2=5.065~$ps, $\phi=-6.33$, $t_0=4.76~$ps, and we also set a weak pump field with strength $e^2A_0^2/(Dm)=0.12/\lambda_s(T=0.8T_c)$. The gap dynamics is then solved via the Runge-Kutta method. We set a damping time $\gamma/D=0.03~$ps during this dynamic simulation.

\section{Formation of   superconducting island during simulation}

\begin{figure}
\centering
\includegraphics[width=15.6cm]{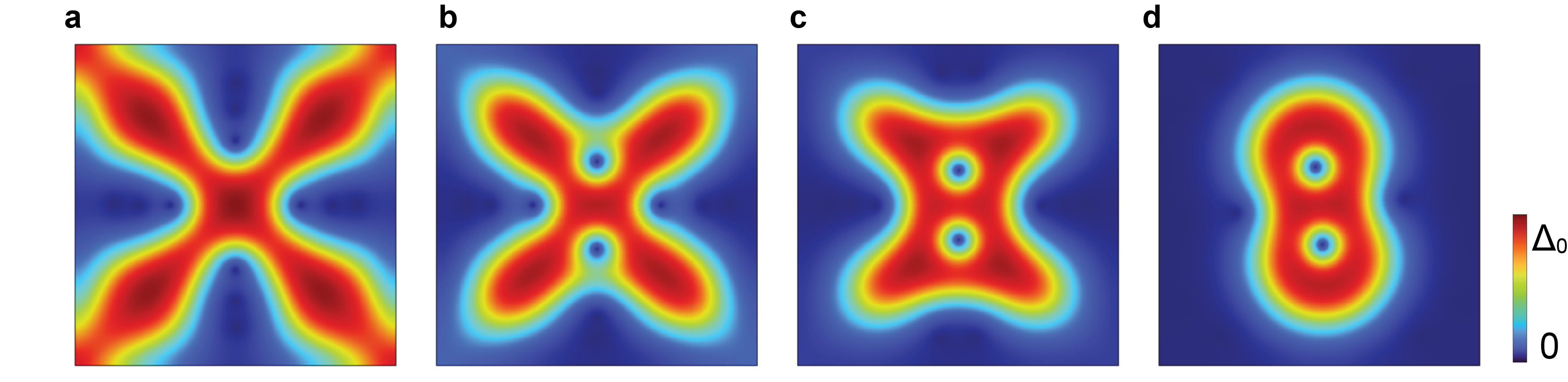}
\caption{Real-time evolution of spontaneous vortex formation corresponding to the
steady-state configuration shown in panel~{\bf h} of Fig.~\ref{fig:pinning}.}
\label{fig:evolution_h}
\end{figure}

Figure~\ref{fig:evolution_h} illustrates the real-time evolution of the superconducting
order-parameter amplitude during the formation of the steady-state configuration at a
high magnetic field, as shown in Fig.~\ref{fig:pinning}(h). Starting from an initially
nearly homogeneous superconducting state, local suppression of $|\Delta(\mathbf{x})|$
first develops near the system boundaries due to the applied magnetic field.

In contrast to the high-temperature regime or low-temperature regime with relatively low magnetic field, where vortices nucleate directly, the
destruction of superconductivity at low temperature with high magnetic field is primarily driven by phase
disordering rather than amplitude suppression. As time evolves, superconducting islands
spontaneously emerge within a background of the normal state. Subsequently, vortices
nucleate inside these superconducting islands and reorganize through dissipative dynamics
until a steady-state configuration is reached.

As mentioned in the main text, this mixed state consisting of superconducting islands is
consistent with previous microscopic calculations~\cite{dubi2007nature}, where the
superconducting transition under strong magnetic fields is driven by phase disordering
and superconducting islands are embedded in a normal-state background. To the best of
our knowledge, such an island-dominated mixed state does not naturally emerge from the
conventional GL formulation, which primarily captures amplitude-driven
suppression of superconductivity near $T_c$.

\bibliographystyle{iopart-num}
\bibliography{ref}

\end{document}